# Enhanced Ferromagnetism in Lacey Reduced Graphene Oxide Nano-ribbon

Vikrant Sahu,[a] V. K. Maurya,[b] S. Patnaik,[*b] Gurmeet Singh[a] and Raj Kishore Sharma[*a]

[a]Department of Chemistry, University of Delhi, New Delhi-110007, India
[b]School of Physical Sciences, Jawaharlal Nehru University, New Delhi-110067, India

## Abstract

Incorporation of magnetism in graphene based compounds holds great promise for potential spintronic applications. By optimizing point defects and high edge density of defects, we report many-fold increase in the ferromagnetic saturation moment in lacey reduced graphene oxide nanoribbons (LRGONR) as compared to other graphene derivatives. The samples were synthesized using chemical unzipping methodology. Detailed structural and morphological characterizations are discussed that include XRD, Raman, SEM, HRTEM and XPS measurements. Brilluoin function analysis to magnetization data reflects best fit for $J = 7/2$ with a saturation moment of 1.1 emu/g. The microscopic origin of magnetization in LRGONR is assigned to high edge defect density which has also been correlated to microstructure.

----------------------------------------------------
Raj Kishore Sharma
*drrajksharma@yahoo.co.in
S. Patnaik
*spatnaikjnu@gmail.com

# 1. Introduction

The prospect of achieving long range ferromagnetic order by tuning defect and edge state morphology of non-magnetic materials, particularly in carbon derived nanomaterials, has attracted considerable attention in the recent past. It promises to revolutionize several aspects of graphene based electronics and spintronics.[1-4] From the synthesis perspective, magnetism in carbon is identifiably linked to the existence of vacancy defect (disorder), reduced dimension, hydrogen chemisorption, grain boundaries etc.[5-9] Leading on these ideas, the evidence for ferromagnetism and anti-ferromagnetism in pyrolytic graphite, nanographites, nanodiamonds and disordered carbon films is now well established.[10-13] With relatively small magnetization (M) (typically, less than ~0.1 emu/g, i.e., less than 0.1% of the magnetization of iron), a consensus has evolved that despite the absence of *d*- or *f*- electrons, magnetism in carbon systems is controllable by synthetic parameters. Towards this end, it is reported that atomic scale defects in graphene-based materials, e.g. adatoms and vacancies, can carry a magnetic moment of over one Bohr magneton, $\mu_B$.[14-18]

In this background, Graphene Nanoribbons (GNRs) with long and reactive edges that are prone to localized electronic states are considered extremely promising towards achieving room temperature ferromagnetism. Moreover, covalent attachment of chemical groups can further alter their electronic and magnetic properties. It is well established that one of the main properties which creates magnetism in graphene is zigzag edges because of the presence of spin-polarized electron state (edge-state) confined in the zigzag edge region. Such prototypical edge morphology has been achieved in potassium-split graphene nanoribbons (PSGNRs), and oxidative unzipped of chemically converted graphene nanoribbons (CCGNRs).[19] In this communication, we report significant enhancement in saturation magnetization by carefully optimizing defect structures in Lacey reduced graphene oxide nanoribbons (LRGONR) prepared by unzipping the multi wall carbon nanotubes (MWCNT). The high ferromagnetism observed is attributed to the existence of high density of state in the LRGONR structures.

## 2. Experimental Section

### 2.1 Synthesis of LGONR and LRGONR samples

Synthesis of lacey graphene oxide nanoribbon and lacey reduced graphene oxide was carried out using chemical unzipping medhodolgy.[20] In this method, 1.5 g of MWCNT was stirred in a mixture of concentrated $H_3PO_4$:$H_2SO_4$ (200 ml) for 15 min. To the stirring mixture, 9g of $KMnO_4$ was added gradually maintaining the temperature of 80 $^o$C for 12 hours. Afterwards the mixture was cooled to room temperature followed by transfer unto a beaker with a mixture of 400ml ice and 5ml $H_2O_2$ (30%). A brown color mixture was formed with slightly higher temperature due to the exothermic reaction that was allowed to cool down to room temperature. Centrifugation of mixture at 8000 rpm was done for 1 hour and brown color precipitate was washed subsequently with 20% HCl, ethanol and DI water to obtain the LGONR. For the synthesis of LRGONR from LGONR, first LGONR was stirred for 24 hrs in a 6 w/w KOH followed by heating in argon atmosphere at 600 $^o$C for 1 hour. To extract the LRGONR from the mixture, KOH was washed away using de-ionized water (till neutral pH was achieved) and the black color powder was dried in vacuum for overnight. Schematic representation of LRGONR synthesis is depicted in Fig. 1.

## 2.2 Materials Characterizations

X-ray diffraction patterns of LGONR, and LRGONR were recorded at X-ray diffractometer (model D8 DISCOVER). Renishaw Invia Reflex Micro-Raman spectrometer was used to record Raman spectra of the LGONR and LRGONR (514 nm wavelength Ar+ laser was used to excite the samples). Perkin-Elmer model 125 used for XPS (X-ray photoelectron spectroscopy) analyses of samples. Phillips Technai T-300 transmission electron microscope and Zeiss Ultra 55 field emission scanning electron microscope were used to capture the HRTEM and FESEM micrographs. The SSA (specific surface area), $N_2$ adsorption-desorption, and pore size distribution were carried on Micromeritics ASAP 2020. Magnetic characterizations were carried on a Vibrating Sample Magnetometer (VSM) attachment in conjunction with *Cryogenic* make Physical Property Measurement System (PPMS). In this paper we report detailed characterizations primarily on LRGONR sample with LGONR sample used as a reference.

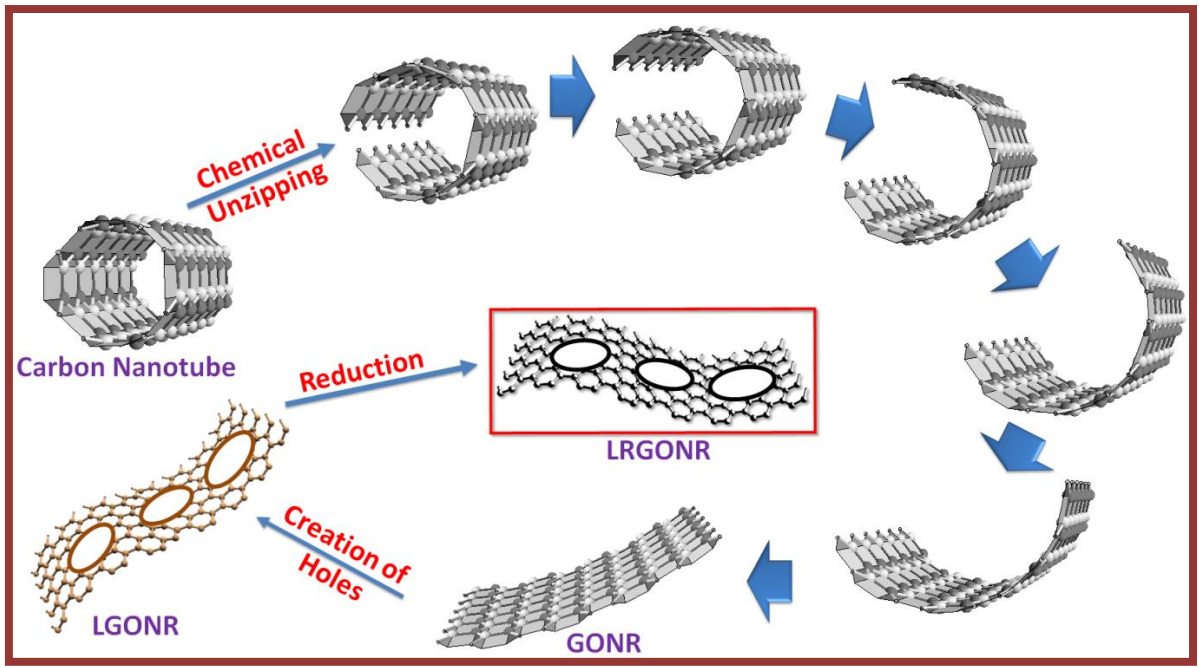

**Fig. 1**. Schematic representation of LRGONR synthesis using MWCNT. Chemical unzipping converted MWCNT into graphene oxide nanoribbon (GONR) and prolonging the process over 12 hours creates holes leading to Lacey graphene oxide nanoribbons. Finally, LGONR was reduced with KOH at high temperature to yield LRGONR.

## 3. Results and Discussion

### 3.1 Structural an d morphological characterizations

Fig. 2a shows SEM micrograph of LRGONR that elucidates the graphene nanoribbons with holes/defects leading to *lacey* patterns. In high resolution transmission electron micrcograph (HRTEM) shown in Fig. 2b, the dimension of holes are estimated to be 30-50 nm. The HRTEM micrograph of a ribbon (Fig. 2b) also shows that layers of GNRs with different width are stacked on one another.

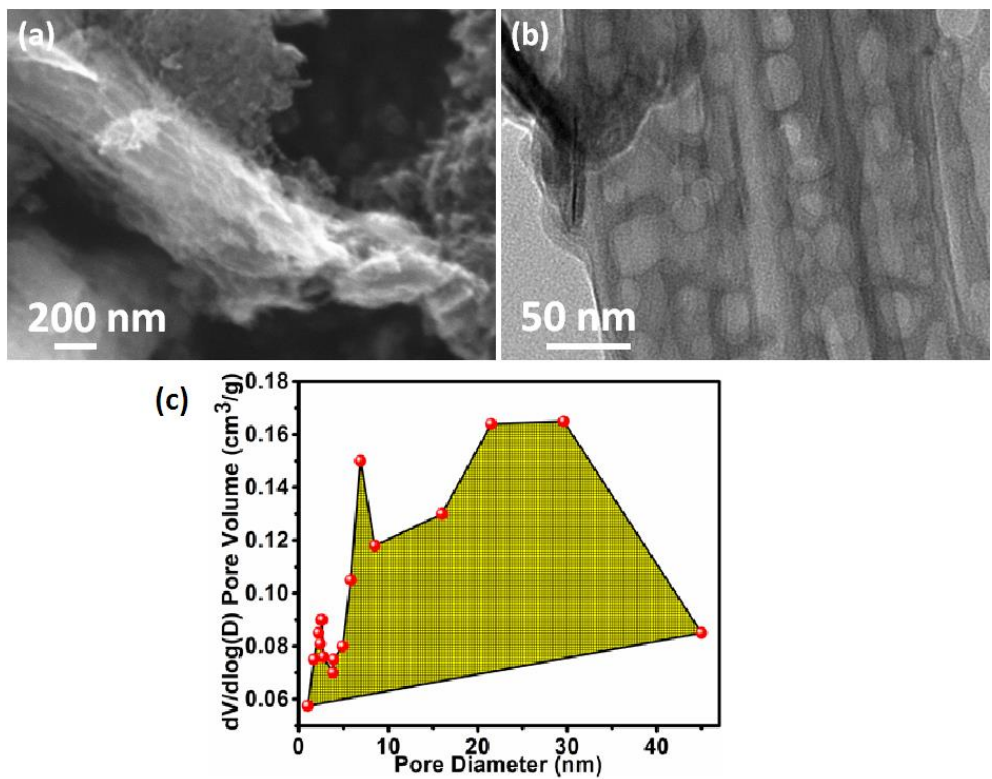

**Fig. 2.** (a) SEM micrograph of LRGONR showing porous stacked nanoribbons, (b) HRTEM micrograph of LRGONR presenting a view of overlapped holey nanoribbons with variable size of defects and (c) plot of pore size distribution of LRGONR.

Evaluation of defects was also carried out through pore size profile obtained from the Brunauer–Emmett–Teller (BET) surface area analysis. Fig. 2c reflects pore volume as a function of pore diameter. Evidently, the maximum contribution to pore volume comes from pores with diameter 30-45 nm which is in qualitative agreement with the HRTEM micrograph.

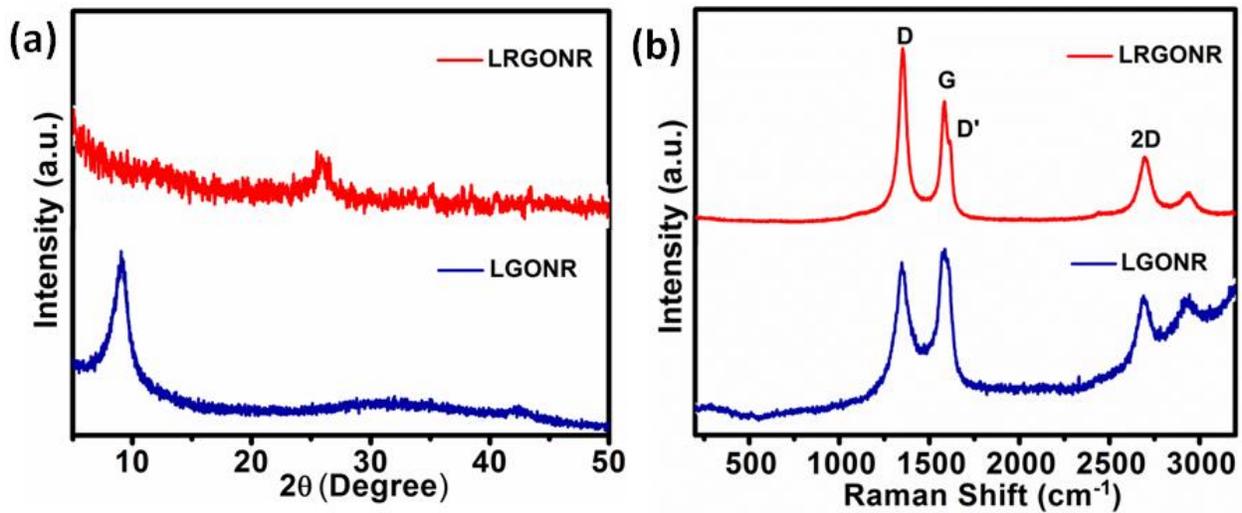

**Fig. 3.** (a) X-ray diffraction pattern of LGONR and LRGONR showing prominent peak of 002 plane at 9.8° and ~25.8°, (b) Raman spectra of LGONR and LRGONR illustrating three major peaks D, G and 2D respectively.

X-ray diffraction pattern of LGONR and LRGONR is shown in Fig. 3a. The peak at 9.8° in LGONR confirms formation of graphene oxide (d-spacing = 9.6Å). With reduction, the peak is expected to shift to higher angles with upper bound of ~26 for pure graphene. The LRGONR sample on the other hand peaks around ~25.8 corresponding to d-spacing of (3.4Å). In comparison the reduced graphene oxide, extracted from MWCNT the d spacing is ~3.4 Å. Since the stacking is low in LRGONR (3-4 layers), a weak peak is observed at ~25.8°. Raman spectra of LGONR and LRGONR are plotted in Fig. 3b. Unambiguous characteristic D and G peaks are observed both in LGONR and LRGONR samples. The intensity ratio ($I_D/I_G$) of D and G peak of LGONR and LRGONR was calculated to be 0.93 and 1.39 respectively. Increase in the intensity ratio of LRGONR is due to the reduction of LGONR, where $sp^2$ carbon clustering and further creation of holes in graphene nanoribbons by KOH leads to reduction at high temperature.[21] The broad 2D peak of LRGONR is positioned at 2697.7 cm$^{-1}$ which is ~19 cm$^{-1}$ in shift compared to the monolayer graphene (2679 cm$^{-1}$) suggesting the 2-4 layer of stacked graphene nanoribbons.[22] Moreover the D' shoulder peak in G band of LRGONR is related to defects in lattice which is correlated to magnetism in LRGONR.[23] The possibility of high density of edge defects is also indicated from the high D peak intensity of LRGONR in comparison to LGONR.[24]

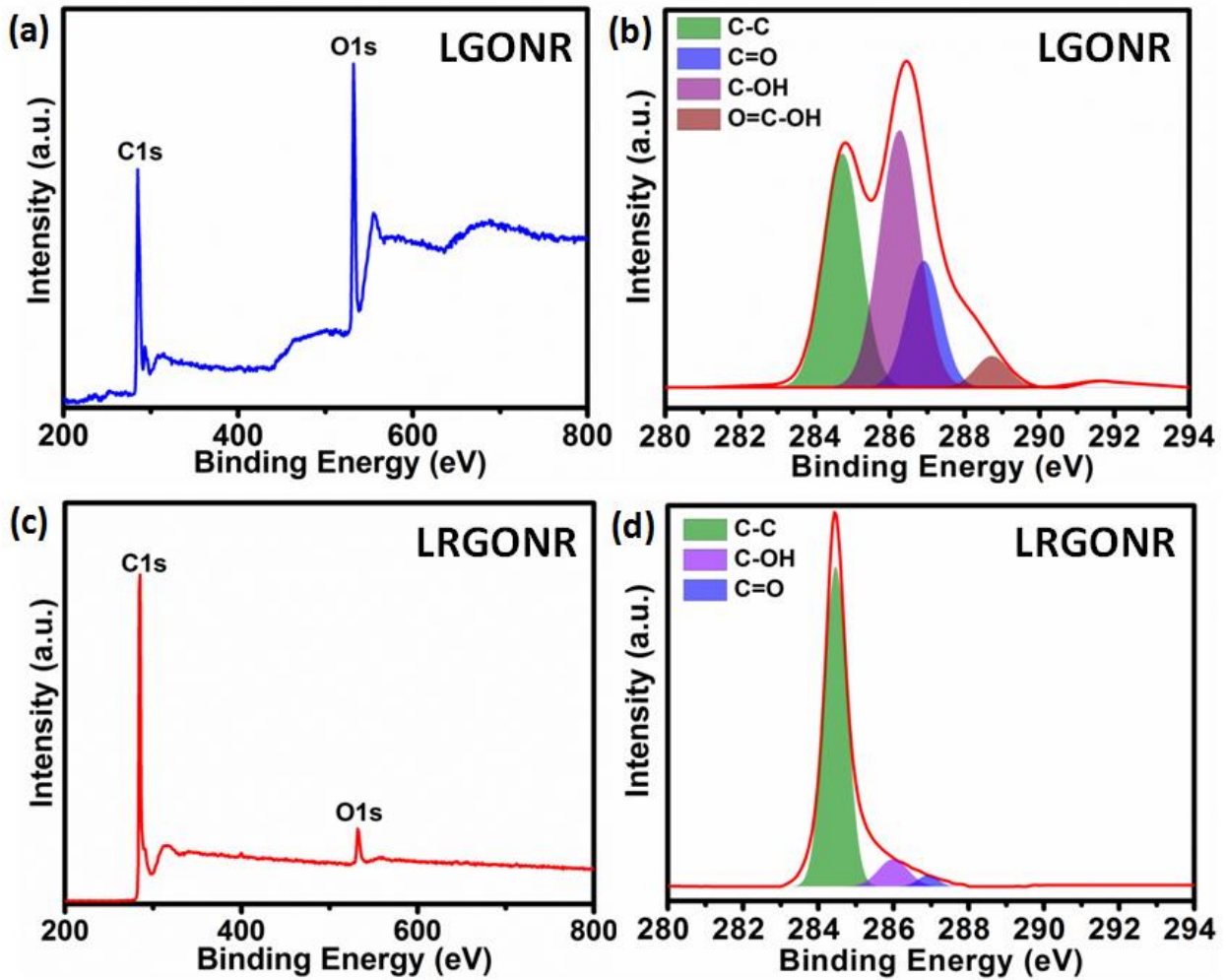

**Fig. 4.** (a) XPS survey spectrum of LGONR illustrating two similar intensity C1s and O1s peaks, (b) C1s core level spectrum of LGONR shows three band of oxygen functionalities with C-C bond, (c) XPS survey spectrum of LRGONR showing two peaks C1s with high intensity and O1s with lower intensity and (d) C1s core level spectrum of LRGONR shows reduced band of two oxygen functionalities with intense peak of C-C.

Towards deciphering oxygen functionalities in LGONR and LRGONR and carbon-oxygen concentration ratio, next we discuss the X-ray Photoemission Spectroscopy (XPS) data. Survey spectra of both LGONR (Fig. 4a) and LRGONR (Fig. 4c) show the two prominent peaks of C1s and O1s but with different intensity. Since LGONR is highly oxidized form of graphene nanoribbons, the peak of O1s is higher in comparison to C1s in LGONR (Fig. 4a). On the other hand, in LRGONR (Fig. 4c) opposite trend is observed due to the reduction. Absence of other peaks in both survey spectra clearly shows that the impurities that can give magnetic properties in MWCNT is washed off during LGONR synthesis procedure. To know the type of C-O functionalities, core level XPS spectra was recorded for both LGONR and LRGONR. Deconvoluted spectra of LGONR (Fig. 4b) show that C-C, C-OH, C=O and O=C-OH functionalities occur at 284.7, 286.2, 286.8 and 288.7 eV respectively.[25] With reduction, the functionalities in LRGONR are excepted to be negligible yet some remnant features are observed (Fig. 4d). However, it is clear that reduction at high temperature resulted in a significant decrease of Carbon/Oxygen ratio from 1.10 (LGONR) to 15.6 (LRGONR). Consequently weak signal of only C=O and C-OH functionalities seen in C1s spectrum of LRGONR.

**3.2 Magnetic characterizations**

In the inset of Fig. 5, room temperature magnetization curve of commercial MWCNT is plotted. A clear hysteretic behavior indicating ferromagnetism is inferred. But this is due to catalyst (derived from Fe/Co/Ni) used during the synthesis of carbon nanotubes. The main panel of Fig. 5 shows the magnetization after turning MWCNT into LGONR. A clear paramagnetic behavior is observed. This implies that the magnetic impurities present in the MWCNT get washed off during chemical unzipping and subsequent washings. The temperature dependent magnetization data for LRGONR is shown in Fig. 6. Zero field cooled and field cooled data show similar trend and are almost superimposed on one another. The data are taken during warming cycle at 0.3 T applied magnetic field. We observe that magnetic susceptibility increases with lowering of temperature down to 30 K and then it increases rapidly and reaches maximum at lowest temperature of 5 K.

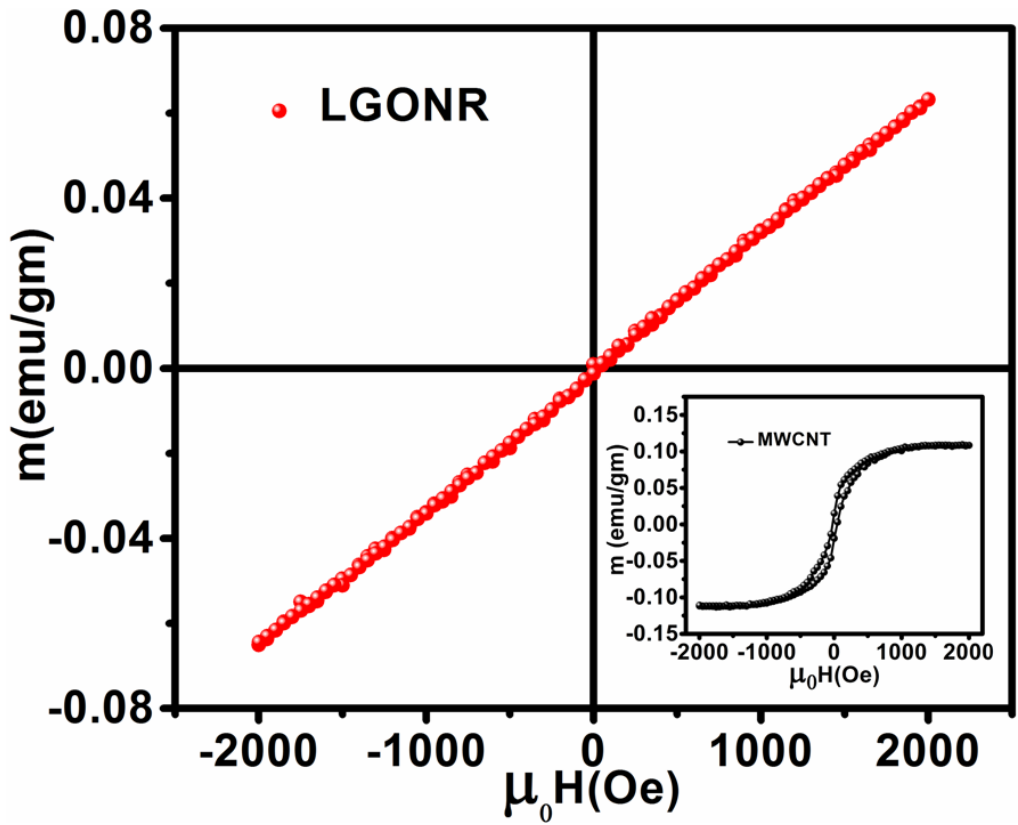

**Fig. 5.** M-H curves LGONR at room temperature. The inset shows hysteretic loops of MWCNT starting material. The ferromagnetism of MWCNT comes from catalysts used during synthesis that gets washed off during the preparation of LGONR.

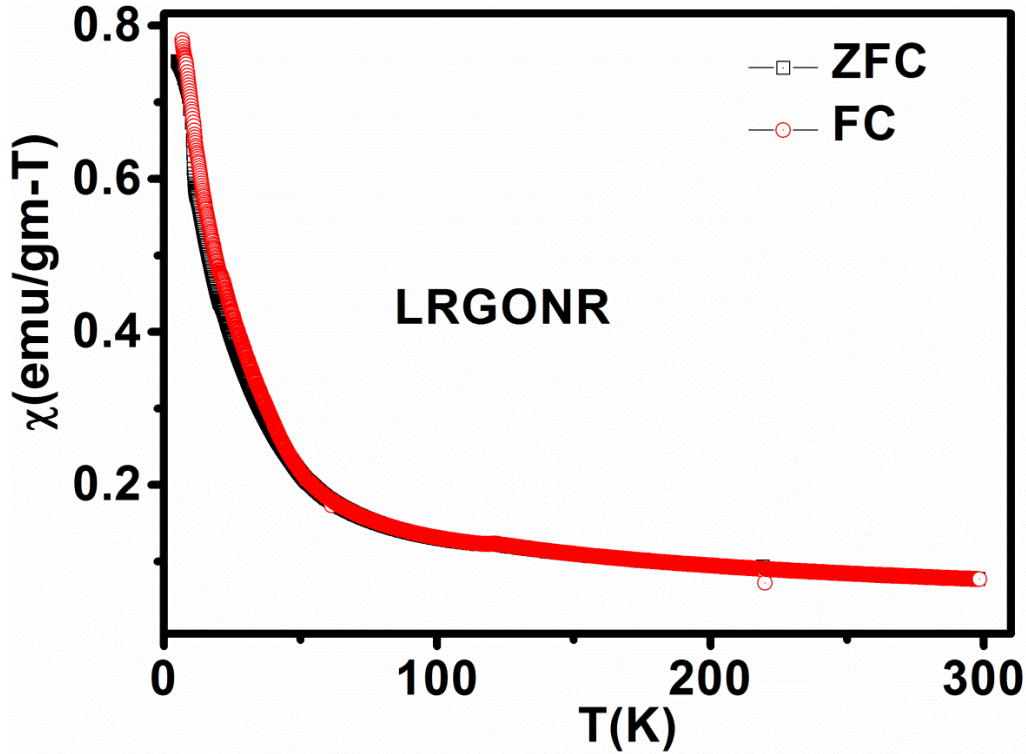

**Fig. 6.** ZFC and FC susceptibility for LRGONR shows sudden increase at 30K. The external magnetic field for this measurement was set at 0.3 Tesla.

To probe the origin of magnetism in LRGONR, we have studied isothermal magnetization measurements on the sample in Fig. 7. Isothermal magnetization curves (M-H) are shown for 2, 50, 100 and 300 K. We can see typical feature for ferromagnetic like behavior at 2 K but at higher temperature hysteresis or non-linear saturation are not seen. The coercivity and remanent magnetization are estimated to be 350 Oe and 0.101 emu/g respectively at 2 K. For comparison, a coercivity value of 160 Oe is reported in nitrogen doped graphene oxide[26] and about 250 Oe is achieved in potassium split graphene nanoribbons.[19] Further, no hysteresis loop is observed at 300 K in Fig. 7.  In a simple two component model, we can assign the magnetization in LRGONR as a sum of two temperature dependent terms; $M_{total} = M_{para} + M_{ferro}$. Evidently, the paramagnetic contribution is dominant at high fields (see inset of Fig. 7) that leads to non-saturated magnetization. In the inset of Fig. 7 we have shown the magnetization (M) data with increasing field fitted to the Brillouin function (Equation 1) with a linear correction term;

$$M = M_0 \left[\frac{2J+1}{2J}\coth\left(\frac{2J+1}{2J}x\right) - \frac{1}{2J}\coth\left(\frac{x}{2J}\right)\right] + M_1 H \qquad \text{Eq.1}$$

Where $x = gj\mu_B H/k_B T$ and M is the magnetization, $M_0$ is $Ngj\mu_B$, $M_1$ is the paramagnetic linear contribution, $\mu_0$ is permittivity of vacuum, $\mu_B$ is the Bohr magneton, $k_B = 1.381 \times 10^{-23}$ J/K is the Boltzmann constant, T = 2 K is the temperature, g is the Lande's g factor which is taken to be 2 and $\mu_0 H$ is the magnetic field in Tesla. Value of J can be integer or half multiples of integers as 1/2, 1, 3/2… and so on which is in the corresponds of magnetically coupled unpaired electrons.[18,27,28] The value of $M_0$ is set to be 0.34 emu/gm and J = 7/2 to fit the experimental data with Brilliouin function. The value of N = no. of effective spins, calculated from $M_0 = Ngj\mu_B$ is estimated to be $5.34 \times 10^{18}$ gm$^{-1}$. The open circles in the inset of graph 6 represent the experimental data at 5 K while the blue curve represents the paramagnetic contribution and red curve which we get the after the subtraction of paramagnetic contribution to the experimental data represents the pure ferromagnetic contribution to total the magnetic moment. The value of saturation moment from the curve is found to be 1.1 emu/gm while Chen et al reported value of saturation moment of ferromagnetic signals at 0.27 emu/gm at 2 K by doping nitrogen in graphene oxides.[26] In similar report a value of saturation magnetization of 0.25 emu/gm is reported at 5 K in potassium split nano-ribbons of graphene by Liu et al.[19]

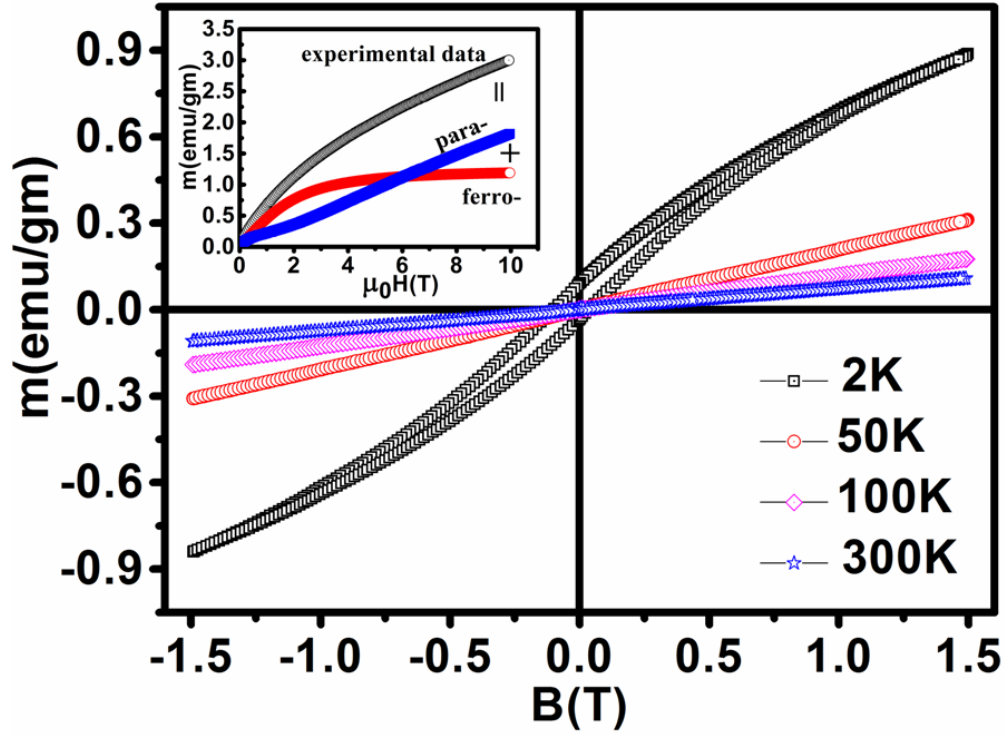

**Fig. 7:** Isothermal magnetic data for 2, 50, 100 and 300 K for LRGONR sample. Sample shows open hysteresis loop at 2 K indicating ferromagnetic behavior. At higher temperature, including at room temperature, sample shows linear M- H behavior. In the inset black open circles represent the experimental data and blue curve represents the calculated para magnetic-contribution of the magnetic moments while red curve represents the ferromagnetic contribution of the magnetic moments after subtracting the paramagnetic contribution.

Next we turn our attention to possibility of exchange –bias phenomena in LRGONR. The field cooled data for M-H loops are shown in Fig. 8. After field cooling, M-H isothermal curves at 2 K temperature shows a shift toward negative field axis at different magnetic fields (0, 0.1, 0.15 and 0.2 T). This phenomenon is reflection of negative exchange bias effect (NEBE). This shift is of common occurrence in a co-existing multi-phase magnetic system. We have studied the cooling field dependence ($H_{FC}$) of exchange bias field ($H_{EB}$) on nano-ribbons. Where $H_{FC}$ is the cooling field, $H_{EB} = -(H_{right} + H_{left})/2$ where $H_{right}$ and $H_{left}$ are the point where M-H loop cut the positive and negative field axis respectively. This study probes the intrinsic magnetic behavior of the ribbons. In the inset of this Fig. 8 we have shown $H_{EB}$ vs. cooling field behavior. $H_{EB}$ increasing rapidly as the function of cooling field which is in the accordance of the fact that field cooling ensures a preferential direction in which magnetic moment freezes at 2 K. It increases the exchange anisotropy in the FM –PM interface and a high increase in $H_{EB}$ is observed. We have also plotted the behavior of $H_c$ (coercivity) and $M_r$ (retentivity) versus the cooling field in the inset of Fig. 8. Both are increasing rapidly in accordance with $H_{CB}$.

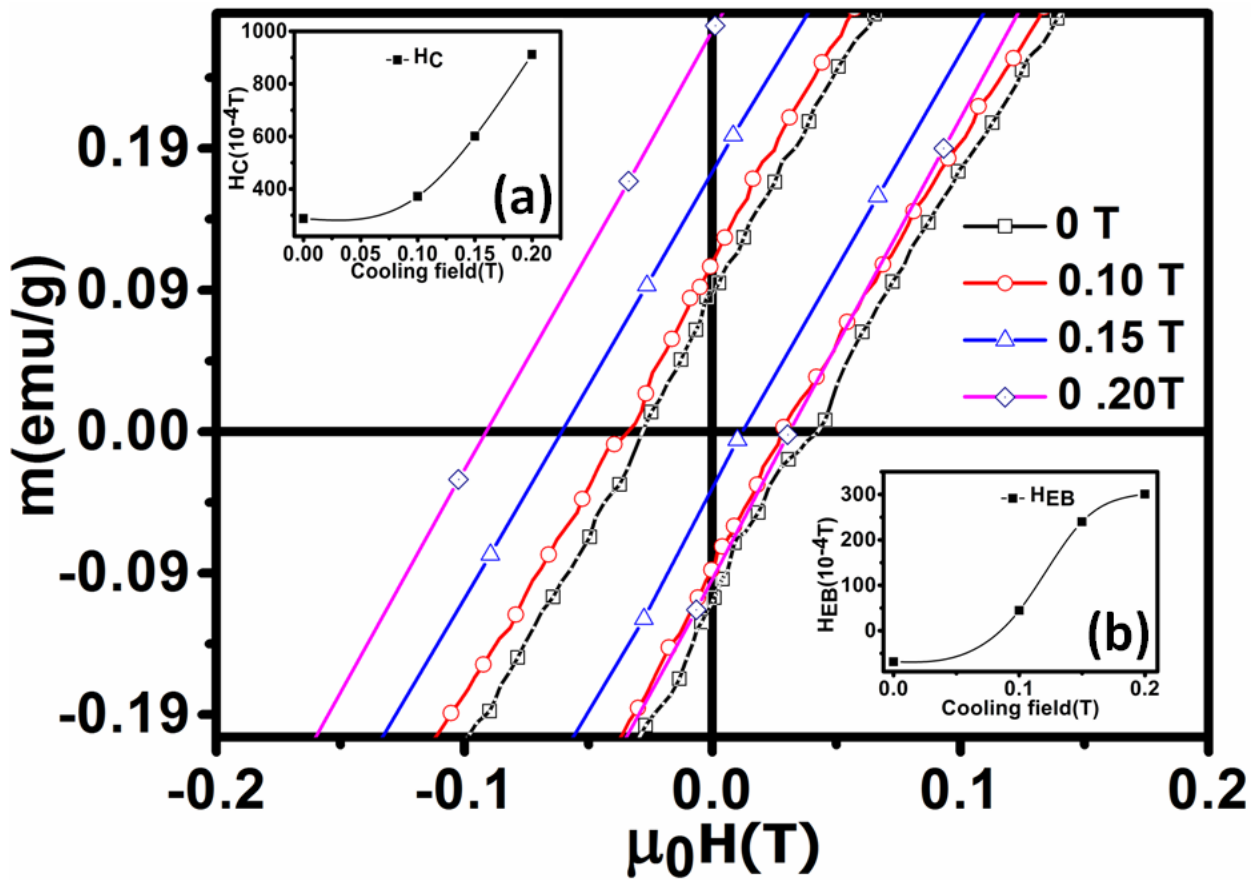

**Fig. 8:** Magnified M-H loops shows negative exchange bias effect in LRGONR sample. Shifting of M-H loop towards negative x-axis of field shows a common feature of co-existing of two different magnetic phases. Coercivity found to be increasing with respect to the cooling magnetic field (inset a). Exchange bias field $H_{EB}$ is also increasing with respect to the cooling field given in (inset b).

**4. Conclusions**

In conclusion, over four fold increase in saturation magnetization is achieved in lacey reduced graphene nanoribbons in comparison to nitrogen doped graphene oxide and similar compositions by optimizing density of zig-zag edge defects. Negative exchange bias effect is observed in LRGNOR which is observed to vary linearly with respect to the increasing magnetic field. The microscopic origin of such enhanced magnetism is assigned to large number of surface disordered Zig-Zag edge defects that are associated with high value of J =7/2.

**Acknowledgments**

The authors gratefully acknowledge University of Delhi for supporting the research through R&D grant (2015-16). Financial support from DST through sponsored research (SR/S1/PC-31/2010) is gratefully acknowledged. One of the authors, V. S. especially acknowledges the SRF award (9/45(1269)/2013-EMR-I) from CSIR, INDIA.